\documentclass[twocolumn,reprint,aps,prd,nofootinbib,floatfix,superscriptaddress]{revtex4-1}
 \usepackage[unicode]{hyperref} 
\usepackage[utf8]{inputenc}
\usepackage{amsmath}
\usepackage{amsfonts}
\usepackage{amssymb}
\usepackage[left=2cm,right=2cm,top=2cm,bottom=2cm]{geometry}
\usepackage{braket}
\usepackage{dsfont}
\usepackage{hyperref}

\usepackage{graphicx}
\usepackage{tikz}
\usetikzlibrary{arrows,decorations.pathmorphing,backgrounds,positioning,fit,petri,shapes}
\usepackage{lipsum}

\usepackage{multirow}
\begin{document}

\title{Unsupervised Searches for Cosmological Parity-Violation: An Investigation with Convolutional Neural Networks}

\begin{abstract}
 Recent measurements of the $4$-point correlation functions (4PCF) from spectroscopic surveys provide evidence for parity-violations in the large-scale structure of the Universe. If physical in origin, this could point to exotic physics during the epoch of inflation. However, searching for parity-violations in the 4PCF signal relies on a large suite of simulations to perform a rank test, or an accurate model of the 4PCF covariance to claim a detection, and this approach is incapable of extracting parity information from the higher-order $N$-point functions. In this work we present an unsupervised method which overcomes these issues, before demonstrating the approach is capable of detecting parity-violations in a few toy models using convolutional neural networks.  This technique is complementary to the 4-point method and could be used to discover parity-violations in several upcoming  surveys including DESI, Euclid and Roman.

\end{abstract}

\author{Peter L.~Taylor}
\email{taylor.4264@osu.edu}
\affiliation{Center for Cosmology and AstroParticle Physics (CCAPP), The Ohio State University, Columbus, OH
43210, USA}
\affiliation{Department of Physics, The Ohio State University, Columbus, OH 43210, USA}
\affiliation{Department of Astronomy, The Ohio State University, Columbus, OH 43210, USA}
\author{Matthew Craigie}
\affiliation{School of Mathematics and Physics, The University of Queensland, QLD 4072, Australia}
\author{Yuan-Sen Ting} 
\affiliation{Research School of Astronomy \& Astrophysics, Australian National University, Cotter Rd., Weston, ACT 2611, Australia}
\affiliation{School of Computing, Australian National University, Acton, ACT 2601, Australia}
\affiliation{Department of Astronomy, The Ohio State University, Columbus, OH 43210, USA}
\affiliation{Center for Cosmology and AstroParticle Physics (CCAPP), The Ohio State University, Columbus, OH
43210, USA}

\maketitle

\section{Introduction}
A detection of parity-violation in the large scale structure of the Universe would break the standard cosmological model. Such a measurement could provide crucial insights into the physics of the early Universe~\cite{Grasso:2000wj, Planck:2015zrl, Shiraishi:2012sn, Cook:2011hg, Adshead:2015pva, Cabass:2022oap, Jazayeri:2023kji} or exotic physics~\cite{Lue:1998mq, Bartolo:2017szm, Alexander:2009tp, Alexander:2009tp, Bartolo:2014hwa, Bordin:2020eui, Pogosian:2007gi, Rybak:2021scp, Ozsoy:2021onx}, and parity-violations may even be responsible for the observed matter-antimatter asymmetry~\cite{Schmidt:2015xka}.
\par Searches for parity-violations have concentrated on the polarized cosmic microwave background (CMB) signal~\cite{Lue:1998mq, Kamionkowski:2010rb, Shiraishi:2010kd, Minami:2020odp,Philcox:2023ffy, Philcox:2023ypl}, gravitational waves~\cite{Yunes:2010yf, Wang:2012fi, Zhu:2013fja, Nishizawa:2018srh} and most recently the clustering of galaxies in the late Universe~\cite{Cahn:2021ltp, Hou:2022wfj, Philcox:2022hkh}, which is the primary focus of this work.
\par Several of these studies have found tantalizing hints of parity-violation. Two recent CMB $EB$-measurements report detections at the $2.4\sigma$~\cite{Minami:2020odp} and $3.6\sigma$-level~\cite{Eskilt:2022cff}, although it has been suggested this could be due to dust emission (see e.g. the discussions in~\cite{Clark:2021kze,Eskilt:2022wav,Diego-Palazuelos:2022dsq}).
\par More recently two studies have reported detections of parity-violations in the Baryon Oscillation Spectroscopic Survey (BOSS)~\cite{alam2015eleventh} CMASS sample at the $2.9 \sigma$~\cite{Philcox:2022hkh} (hereafter Ph22), $7.1 \sigma$~\cite{Hou:2022wfj} (hereafter Hou22) and $8.1 \sigma$-levels \cite{Philcox:2021hbm} (hereafter Ph21).  These detections are made from measurements of the 4PCF which are enabled by recent computational advances~\cite{Slepian:2015qza, Slepian:2015qwa, Philcox:2021bwo}. If physical in origin, the cause of this signal must also be consistent with the null detection of parity-violation in the temperature trispectrum of the CMB~\cite{Philcox:2023ffy}.
\par
\begin{figure}[!hbt]
\includegraphics[width = \linewidth]{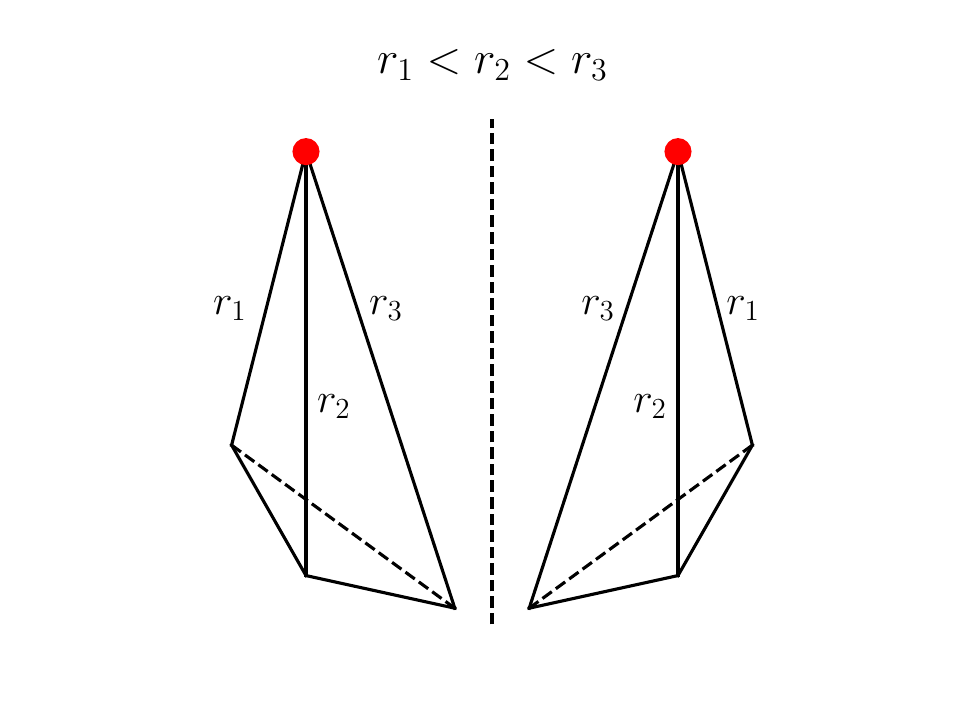}
\includegraphics[width = \linewidth]{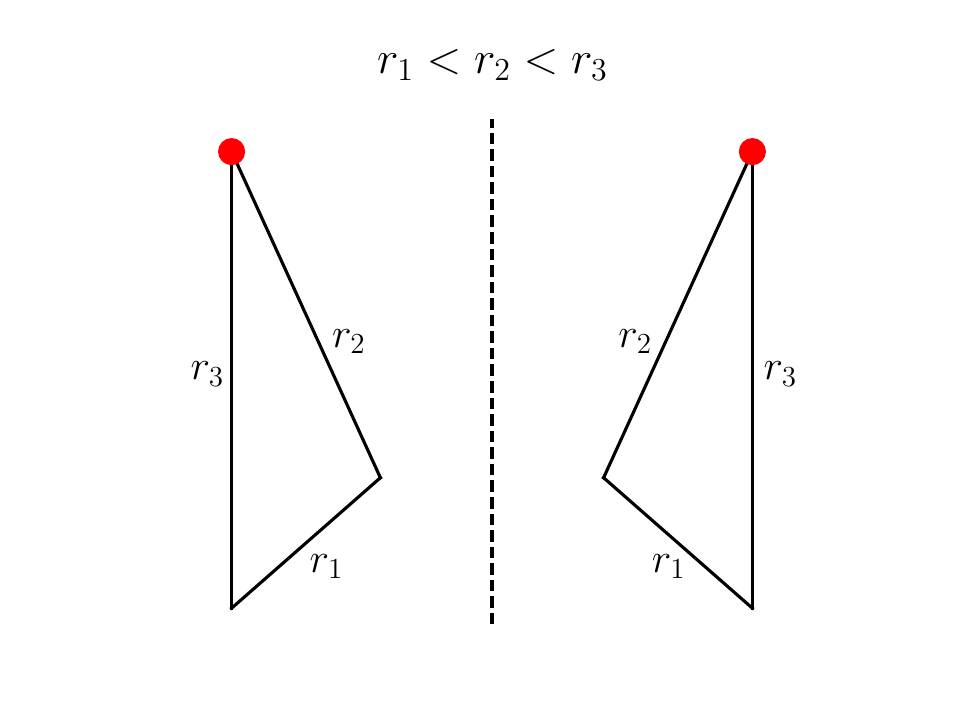}
\caption{ In three dimensions, a tetrahedron is the lowest order polyhedron which is sensitive to parity i.e., handedness.  {\bf Top:} For the purposes of this work we say a tetradedron is right-handed if when moving the right hand to the primary vertex with the thumb pointed out the fingers curl from the shortest to longest edges connected to the primary. It is impossible to rotation a left-handed tetraderon (left) onto it's mirror image (right), which is equivalent to a parity transform up to a $180^\circ$ rotation. {\bf Bottom:} In contrast the left-handed triangle (left) is equivalent to  rotating the right-handed triangle (right) on the right `out of the page'.}
\label{fig:parity int}
\end{figure}

 In three dimensions the 4PCF is the lowest order $N$-point correlation function which is sensitive to parity-violations~\cite{Cahn:2021ltp}. To see this, it is useful to refer to Fig~\ref{fig:parity int} where galaxies are treated as vertices of a tetrahedra. After defining a notion of the tetrahedra handedness about a prespecified primary vertex, we see that mirroring a tetrahedron about a vertex, which is equivalent to a parity transformation up to a rotation, switches the handedness of the tetrahedron. This is in contrast to a triangle or a line segment, where a reflection about a point is equivalent to a three-dimensional rotation. 

\par The 4PCF approach has three main drawbacks:
\begin{itemize}
    \item {First, even in the absence of observational systematics, quantifying the statistical significance is theoretically and computationally demanding. Hou22 quantified the uncertainty by computing the covariance of the 4PCF. The covariance has both a large number of terms and potentially poorly understood 8-point function contributions. This approach also assumes that the parity-odd 4-point function is Gaussian distributed. Meanwhile, Ph22 performed a rank test and compared the level of parity-violation in the data to a suite of 2048 MultiDark-Patchy mock realizations~\cite{Kitaura:2015uqa, Rodriguez-Torres:2015vqa} of the CMASS footprint. These simulations have not been extensively verified at the 4-point level, and may lead to an underestimate of the uncertainty.} 
    \item {Second, while it is in principle possible to extract parity information from the higher order $N$-point functions~\cite{Cahn:2020axu}, this quickly becomes intractable for larger $N$. This potentially leaves a large-part of the parity-odd signal untapped.}
    \item {Since the 4PCF is a global measure, it is difficult to localize which physical regions contribute most to the parity asymmetry. Ideally one could identify the regions where parity-violations are strongest and perhaps even cross-correlate with known systematics to determine whether systematics are responsible for observed parity-violations.}
\end{itemize}
\par To address these shortcomings, we propose an unsupervised method which can in principle detect parity-violation using information from all $N$-point functions {\it from the data alone}. This technique may be employed to search for parity-violations in large spectroscopic surveys including Euclid\footnote{\url{http://euclid-ec.org}}~\cite{Blanchard:2019oqi,Laureijs:2011gra}, the Nancy Grace Roman Space Telescope\footnote{\url{https://www.nasa.gov/roman}}~\cite{spergel2015wide}, the Dark Energy Spectroscopic Instrument (DESI)\footnote{\url{https://www.desi.lbl.gov/}}~\cite{Aghamousa:2016zmz}. 
\par Intuitively, the algorithm described in this paper is simple. The first step is to divide the large-scale structure map into a large number of sub-volumes. Then, for each sub-volume, one can parity-flip the volume before passing both the original and parity-flipped sub-volumes through a convolutional neural network (CNN). If the CNN can reliably classify which is the true realization and which has been parity-flipped, this implies that the data is not parity invariant.\footnote{In practice, the problem is recast as a maximization problem rather than a classification problem. See Sect.~\ref{sec:alg}.}
\par Similar techniques have been employed to determine whether a photograph has been mirrored~\cite{lin2020visual}, distinguish images of mirrors from transparent glass~\cite{tamura2022distinguishing} and classify chirality in medical imaging~\cite{kang2022asymmetry}.  In high-energy physics, a similar technique has been developed to search  the Large Hadron Collider's calorimeter data for evidence of parity-violation~\cite{Lester:2021aks, Tombs:2021wae}. It is the technique of these former two works which this paper follows most closely.  While we focus exclusively on CNNs in this paper, the method is generalizable to any architecture. In a follow up paper (Craigie et. al. in prep), we also consider neural fields and scattering networks and demonstrate that these can significantly outperform CNNs.
\par The unsupervised parity detection algorithm is described in Sect.~\ref{sec:method}. It is then applied to 2D and 3D toy models in Sect.~\ref{sec:method}. Finally in~\ref{sec:conclusion} we conclude, by outlining a road-map for future work.

\section{Unsupervised Searches for parity-violations Using Convolutional Neural Networks} \label{sec:method}

\begin{figure*}[!hbt]
\includegraphics[width = 0.9\linewidth]{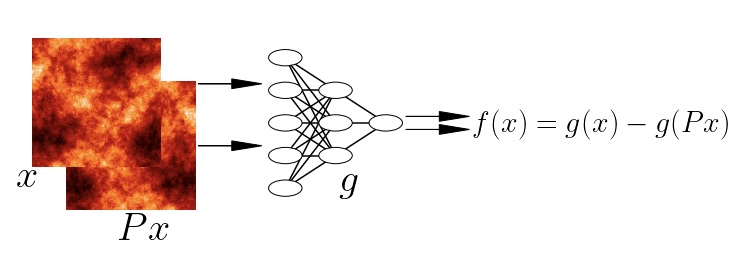}
\caption{Schematic showing the algorithm presented in Sect.~\ref{sec:alg}. Some data, $x$, and the parity-flipped data, $Px$, are both passed through a CNN denoted, $g$, such that $g(x) \in  \mathbb{R}$. For every pair we define, $f(x) = g(x) - g(Px)$. The network learns to maximize the difference between $g(x)$ and $g(Px)$ during training (see  Sect.~\ref{sec:alg} for more details). As shown in Sect.~\ref{sec:detection}, the expected value of $f(x)$ is exactly zero if no parity-violations are present. Any deviation from zero is indicative of parity-violation, while the statistical significance can be estimated from boostrapping over the validation set.}
\label{fig:schematic}
\end{figure*}

\subsection{Algorithm} \label{sec:alg}
The algorithm described below and summarized in Fig.~\ref{fig:schematic} was originally proposed in~\cite{Lester:2021aks} (hereafter Le21) in the context of particle physics. 
\par Consider a CNN, $g$, which maps some input, $x$, to $g(x) \in \mathbb{R}$. In the context of this work, one should think of $x$ as a sub-volume of a pixelized (or voxelized) galaxy over-density map. Let us also define a parity operator, $P$. In two dimensions, $P$ is equivalent to mirror symmetry, while in three-dimensions the parity operator flips a volume along all axes.
\par Intuitively, to detect a parity-violation we would like the network, $g$, to maximize the distance between $g(x)$ and $g(Px)$. To build such a network, define a new function
\begin{equation}
f(x) = g(x) - g(Px).
\end{equation}
Then for a batch, $B = \{ x_1 , x_2 .., x_N\}$, one might na\"ively expect that choosing a loss function to maximize the mean output of the batch would achieve the objective. However, trivially rescaling all the weights and biases of the network by some constant $\alpha > 1$ achieves a smaller loss, so instead we define the loss-function,
\begin{equation} \label{eq:loss}
    -\mathcal{L}_B = \mu_B / \sigma_B,
\end{equation}
where the mean, $\mu_B$, and variance, $\sigma_B$, over the batch are defined as
\begin{equation}
\mu_B = \frac{1}{N} \sum_{x \in {B}} f(x),
\end{equation}
and
\begin{equation}
\sigma_B ^2 = \frac{1}{N} \left( \sum_{x \in {B}} \left[f\left(x\right)\right] ^2 \right) - \mu_B ^2.
\end{equation}
\par After splitting the data into training and validation sets, the network can be trained in the usual fashion before using the validation set to detect parity-violations. We describe this procedure in detail in the following section.

\subsection{Network Behavior, Robustness to Overtraining and Parity-Violation Detection Criteria } \label{sec:detection}
 \par Given the choice of loss in Eqn.~\ref{eq:loss}, the network learns to return a maximal positive number with small variance over the training set. Thus, if the mean value of $f(x)$ over the validation set, $\langle f(x)\rangle_{x \in V}$ (hereafter written $\langle f\rangle_{V}$), is greater than zero, then we have detected parity-violation and the statistical significance can be quantified by bootstrapping over the validation set.
Alternatively one could use $\langle f \rangle_V$ as an extremely compressed parity-odd summary and perform a rank test against a large number of simulations as in Ph21.
 \par Our technique does not make false detections even when the network is overtrained. To see this it is useful to consider the case that no parity-violations are present in the data. Then the probability that any given $x$ is in the validation set is the same as the probability that $Px$ is in the validation set. This observation has an important consequence. Let us suppose that after training $f(x) = n$, for some positive number $n$, then by design, $f(Px) = -n$. This implies that if no parity-violations are present in the data, the expected mean value of  $f(x)$ over the validation set, $\langle f \rangle_{V}$, is precisely zero. As noted in~\cite{Lester:2021aks}, this is true {\it even if the network is overtrained}. 
\par It is worth noting that in order for the bootstrap uncertainties to remain valid, we implicitly assume that there is no covariance between the sub-volumes in the validation set. In practice this can likely be achieved by choosing validation set sub-volumes that are separated by large physical separations in the survey. This must be explicitly checked when working with real data. In (Craigie et. al. in prep), we also explicitly confirm that the distribution of the bootstrap resampled means is the same as the likelihood computed from simulations of independent universes.


\section{Results} \label{sec:results}

\begin{figure*}[!hbt]
\includegraphics[width = 0.49 \linewidth]{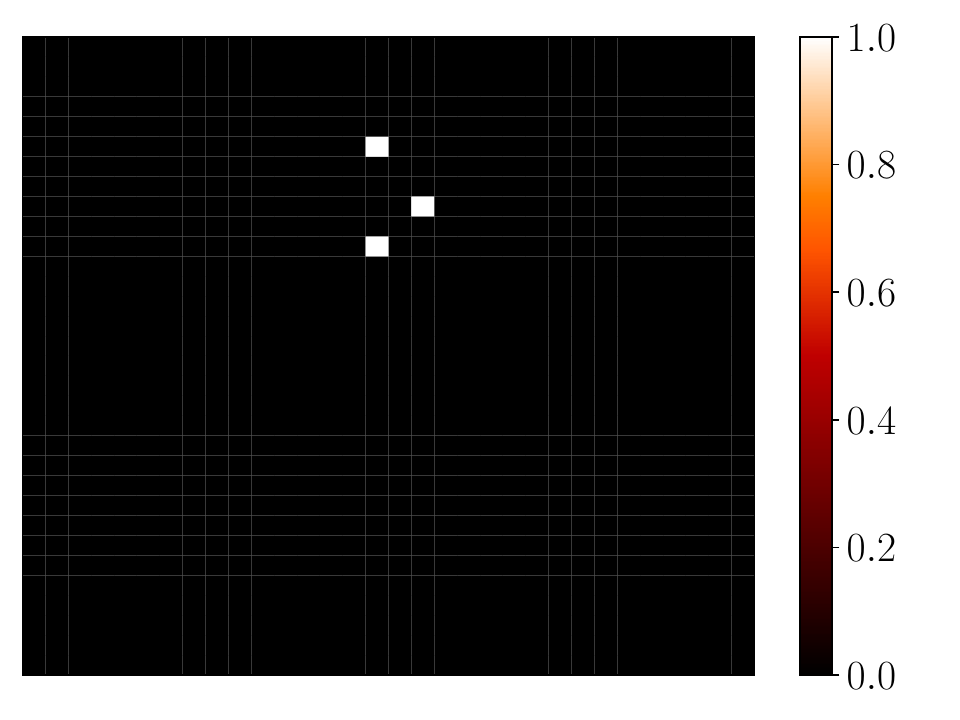}
\includegraphics[width = 0.49 \linewidth]{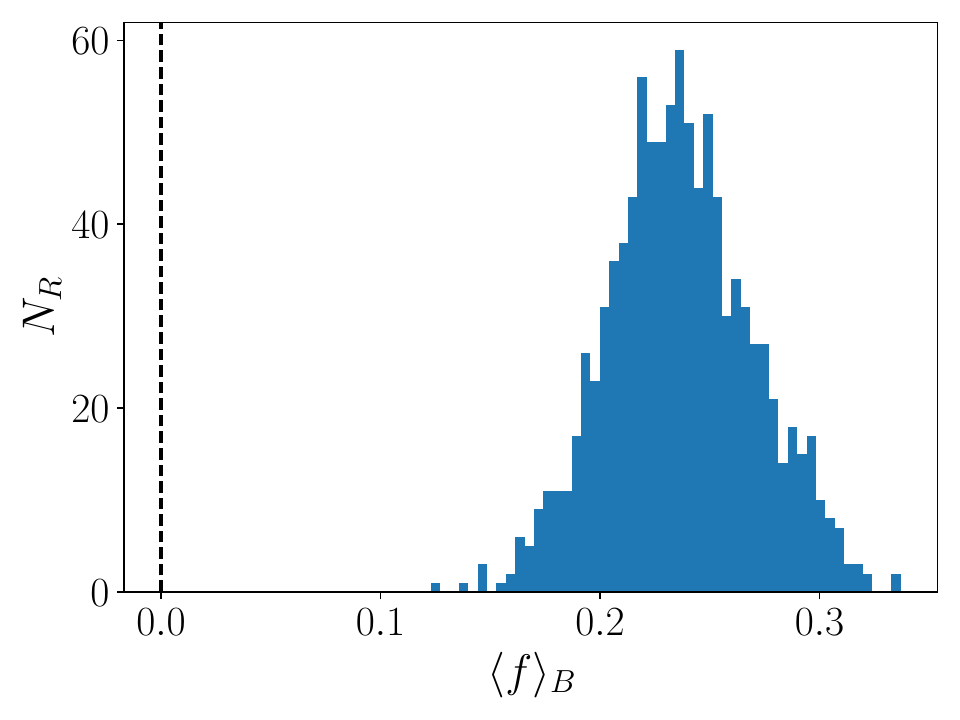}
\caption{{\bf Left:} In this example, parity-violating `fields' are generated by mapping a fixed right-handed triangle, $T$, to a random location in the field and assigning a galaxy to each vertex (see Sect.~\ref{sec:ex1} for more details). {\bf Right:} Histogram of the mean of the validation set over bootstrap re-samplings, $\langle f \rangle _B$, while $N_R$ gives the number of occurrences in the re-sampling in each histogram bin. This is greater than zero (indicated by a dashed line) at high-confidence, so we have detected parity-violations.}
\label{fig:results2d_1}
\end{figure*}

\begin{figure*}[!hbt]
\includegraphics[width = 0.49 \linewidth]{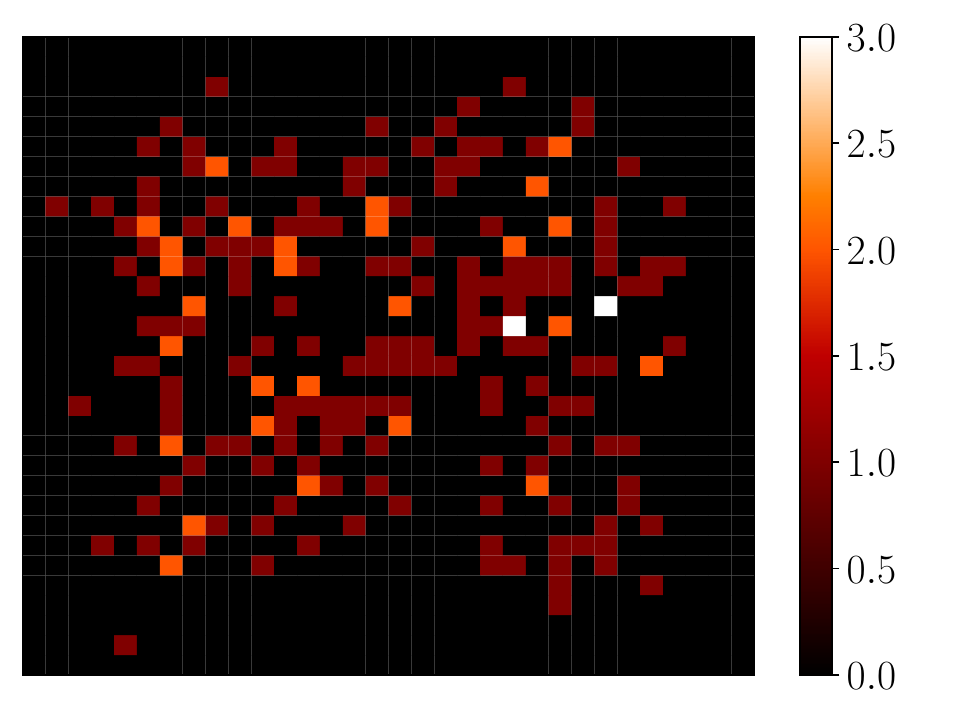}
\includegraphics[width = 0.49 \linewidth]{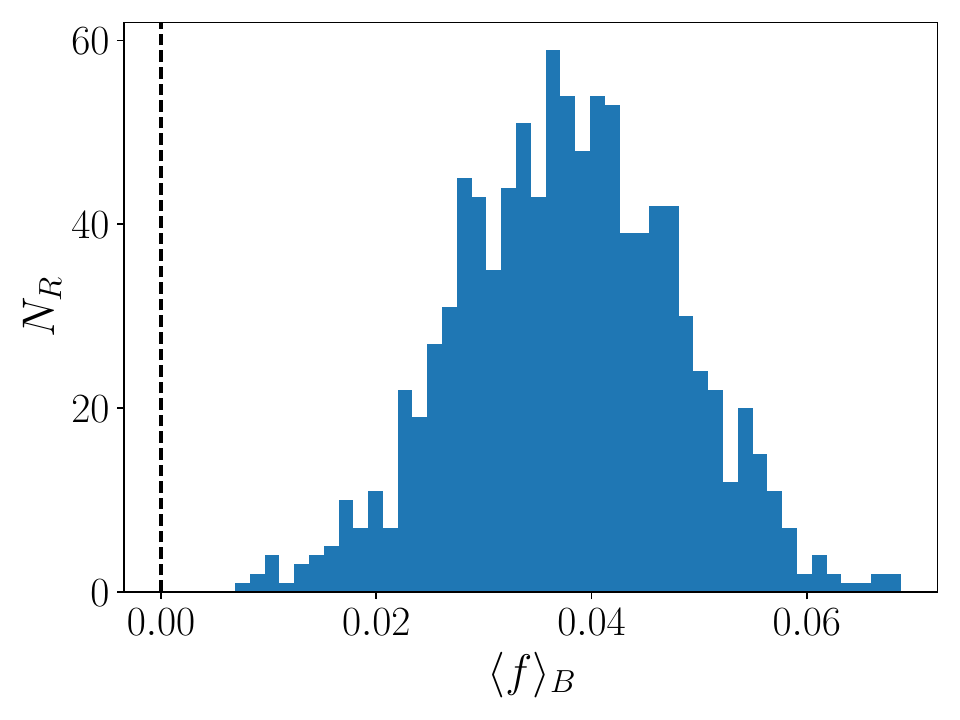}
\includegraphics[width = 0.49 \linewidth]{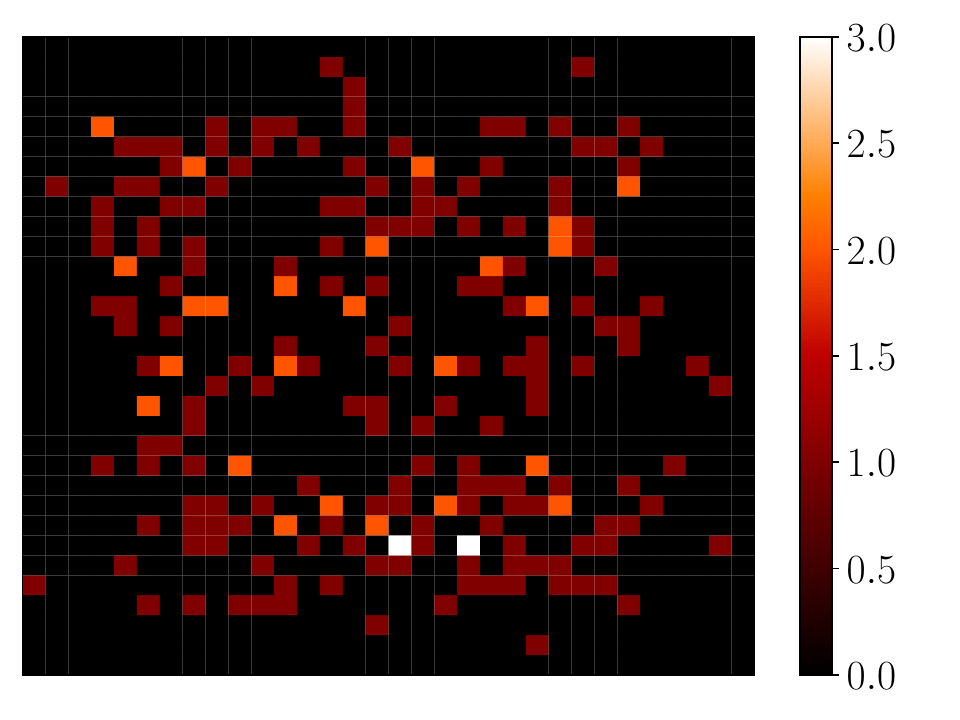}
\includegraphics[width = 0.49 \linewidth]{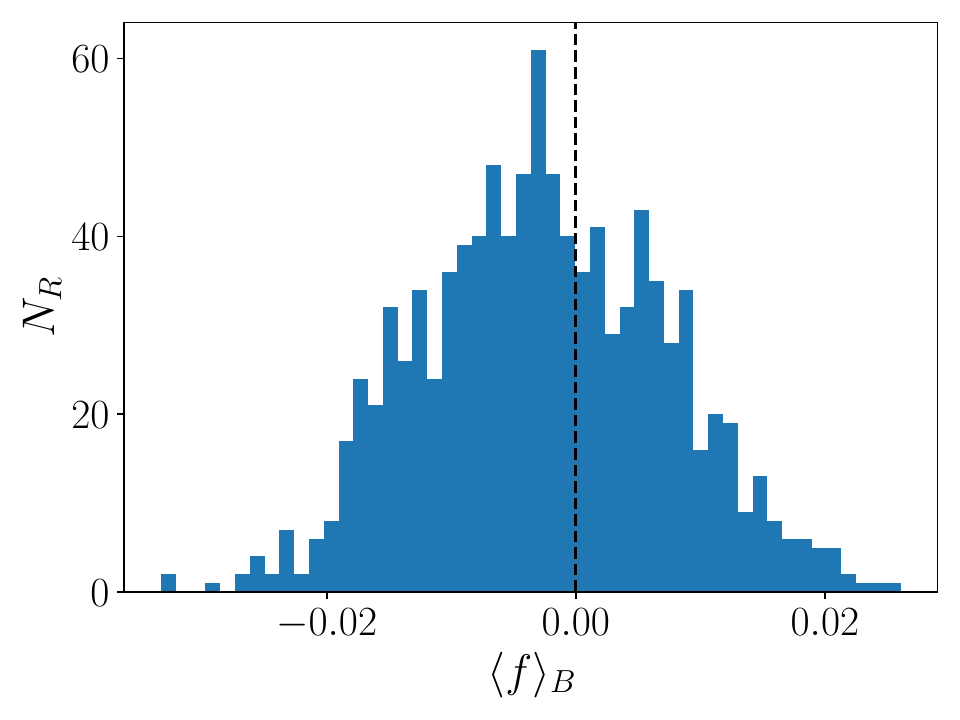}
\caption{{\bf Top left:} An example of a parity-violating `field' generated by mapping left-handed and right-handed triangles (with an overabundance of one kind) to a random location in the field and placing a galaxy at each vertex. The fields have on average 72.5 triangles per field and we place on average 5 more right-handed compared to left-handed triangles per field. Given all the possible triangles that can be formed from these vertices, this represents less than a $1 \%$ relative overabundance of right-handed triangles on average.  See Sect.~\ref{sec:ex2} for more details. {\bf Top right:} Histogram of the mean of $f$ from bootstrap resampling the validation set. Parity-violations are detected at high significance. {\bf Bottom left:} Same as top left, but generated as a null test using  a procedure where there is no overabundance of left versus right handed triangles (see Sect Sect.~\ref{sec:ex2}) for more details). {\bf Bottom Right:} Same as above. No parity-violations are detected in the data as $\langle f \rangle_{B}$ is consistent with zero.} 
\label{fig:results2d_2}
\end{figure*}


\begin{figure}[!hbt]
\includegraphics[width = \linewidth]{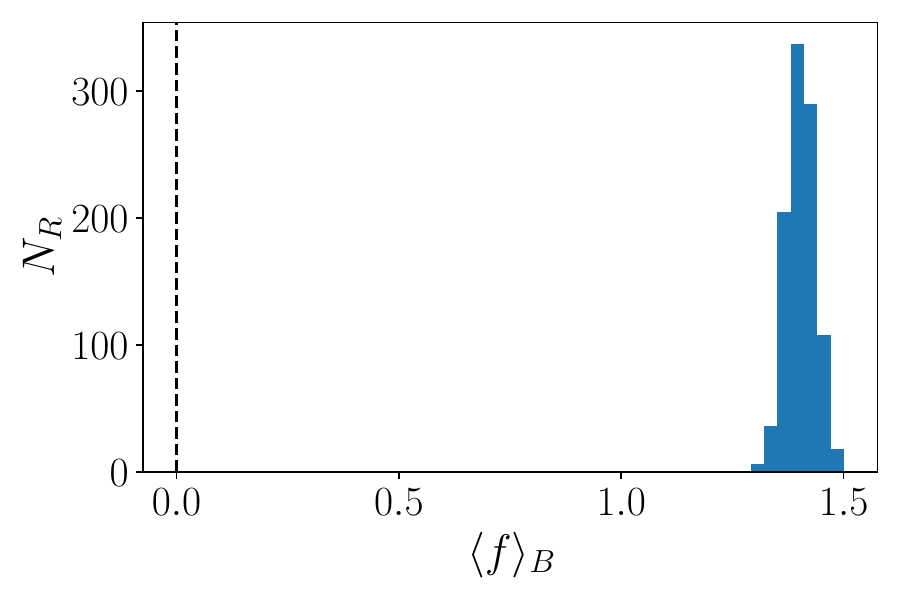}
\caption{Same as the right hand side of Fig.~\ref{fig:results2d_1} but for this example we use a single right-handed tetrahedron per field. We make a clear detection of parity-violation. }
\label{fig:results3d_1}
\end{figure}

In this section, we showcase how the method described in Sect.~\ref{sec:alg} can be used to search for parity-violations in the context of galaxy surveys. In this paper we use a parity-violating toy model described in Sect.~\ref{sec:triangle}. We show how the unsupervised technique is capable of detecting parity-violations in these fields. We  start by building intuition in 2D (see Sect.~\ref{sec:2d}) before demonstrating that the technique extends to 3D (see Sect.~\ref{sec:2d}). All CNNs are implemented in {\tt TensorFlow}~\cite{tensorflow2015-whitepaper} {\tt Keras}~\cite{chollet2015keras}.\footnote{{\tt PyTorch}~\cite{paszke2017automatic} was also used during the development phase.}

\subsection{A Toy Model with Parity and Mirror Asymmetries} \label{sec:triangle}
\par  To generate a 2D mirror asymmetric field, we:
\begin{enumerate}
\item start with a left-handed\footnote{It is equally valid to use a right-handed triangle instead.} triangle (see Fig.~\ref{fig:parity int} for more details) with a vertex placed at the origin,
\item apply a random isometry to the triangle in the plane,
\item place a galaxy inside each voxel containing a vertex of the triangle,
\item and repeat steps 1-3 $N$ times to generate a galaxy field with $3N$ galaxies.
\end{enumerate}
 In 3D, rather than triangles we use right and left-handed tetrahedra. More complex parity-violating fields can be built by mixing right and left-handed tetrahedra or triangles (see e.g. Sect.~\ref{sec:ex2}).

\subsection{Toy Models in Two Dimensions} \label{sec:2d}

\subsubsection{2D Model 1: 1 Triangle Per Field} \label{sec:ex1}
Following the procedure described in Sect.~\ref{sec:triangle}, we use the triangle, $T$, with vertices in the plane at $(0,0)$, $(4,0)$ and $(0,8)$ to generate a set of `random fields'. We apply a random isometry by applying a random rotation, $\theta \in [0, 2 \pi]$, and uniformly drawing a translation, $\Delta x, \Delta y \in [-23,23]$ inside each field\footnote{This choice ensures that all the vertices of the triangle remain inside the field.}. The field is then pixelized on a $32 \times 32$ grid centred on the origin. A randomly chosen realization is shown on the left of Fig.~\ref{fig:results2d_1}.
\par We produce a training set of $10^4$ realizations and a validation set of $10^3$ realizations before training a network to detect parity-violations following the procedure outlined in sect.~\ref{sec:alg}. A full description of the network architecture and training is given in Appendix~\ref{sec:A1}.
\par An example of the training data is shown on the left hand side of Fig.~\ref{fig:results2d_1}. Meanwhile a  histogram of the mean of the validation set over bootstrap samples is shown on the right hand side. This is greater than zero at every resampling so we have made a clear detection of mirror asymmetry. Even by eye, one can easily `detect' mirror asymmetries, so we now consider a more challenging example.

\subsubsection{2D Model 2: Complex Triangle Example} \label{sec:ex2}
 We use the triangle, $T$, defined in the previous example and the mirror image, $PT$. In each field we draw $N_T$ copies of triangle $T$ and $N_{PT}$ copies of triangle $PT$ where $N_T$ and $N_{PT}$ are respectively drawn uniformly from the discrete sets $\{ 31,32,33,34,35,36,37 \}$ and $\{ 42,41,40,39,38,37,36\}$ inducing a small amount of parity-violation. Given all the possible triangles that can be formed from these vertices, this represents less than a $1 \%$ relative overabundance of right-handed triangles on average. We then apply an independent isometry to each galaxy as in the previous example, before placing a galaxy at each vertex and pixelizing. Hence, on average there are 72.5 triangles per field, but with a slightly different amount of mirror asymmetry per field. 
 \par We generate $10^5$ training realizations and $10^4$ realizations for validation so that there are $\sim$$7.5 \times 10 ^6$ `galaxies' in total. This is similar to the expected number of galaxies in upcoming spectroscopic surveys including Euclid, DESI and Roman. 
 \par A randomly chosen example of one of the fields generated following this procedure is shown on the top left of Fig.~\ref{fig:results2d_2}. Unlike the example in the previous section, it is impossible to detect parity-violations by eye. 
 \par To search for parity-violations, we train the network described in Appendix~\ref{sec:A2}. A histogram of the mean over bootstrap resamples of the validation set displayed in the top right of Fig.~\ref{fig:results2d_2}. We clearly detect parity-violations at high significance.
 \par As a null test, we repeat the experiment described above, but this time in each field we draw $N_T$ copies of triangle $T$ and $N_{PT}$ copies of triangle $PT$ from the same discrete set: $\{ 33,34,35,36,37,38,39\}$. With this choice, we expect 72 galaxies per field on average with no overabundance of right-handed or left-handed triangles. 
 \par A randomly chosen example of such a field is shown on the bottom left of Fig.~\ref{fig:results2d_2}. It is virtually indistinguishable from the parity-violating field in the top panel.
 \par We generate a null training set of $10^5$ training realizations and $10^4$ validation realizations before training a network with the same architecture as before. The results are shown on the bottom right of Fig.~\ref{fig:results2d_2}. As expected, the network does not detect parity-violations when they are not present.
 
 \subsection{Toy Models in Three Dimensions} \label{sec:3d}

\subsubsection{3D Model 1: 1 Tetrahedron Per Field} \label{sec:ex4}
In this example we use a tetrahedron, $T$, with vertices at (1,0,0), (-3,3,0), (-5,-5,-5)
and (0,0,3). Following the procedure outlined in~\ref{sec:triangle}, we apply a random rotation defined by 3 Euler angles and a translation $\Delta x,\Delta y,\Delta z\in [-10,10]$. Each vertex is then mapped onto a $28 \times 28 \times 28$-grid before a galaxy is place inside each pixel containing a vertex.
\par We generate $10^4$ training realizations and $10^3$ validation realizations before training the network described in Sect.~\ref{sec:A4} to search for parity-violations. We are able to make a clean detection of parity-violation as shown in Fig.~\ref{fig:results3d_1}. 
\par The purpose of this exercise has been to demonstrate that we can in principle detect parity-violations in 3D. However, in contrast to the 2D case, we find that the CNN fails to extract parity information with a more realistic number of galaxies in each field. In particular given the large number of free parameters in the model, we find that the CNN is prone to over-training and the model does not generalize well to the validation set, which we use for detection. Convolutions in 3D are also prohibitively expensive and inefficient as much of the data volume contains empty pixels. For these reasons we will need to find a more efficient architecture before attempting to detect parity-violations with real data. This is the subject of a follow-up paper (Craigie et. al. in prep).

\section{Conclusion} \label{sec:conclusion}
\par This paper presents a method to perform an unsupervised search for parity-violations from spectroscopic galaxy surveys. This follows possible detections using the 4PCF method in Ph21, Ph22 and Hou22. In principle a similar technique could be used to search for $EB$ parity-violations in weak lensing (see e.g.~\cite{Philcox:2023uor}) or CMB data. 
\par We have demonstrated that CNNs are readily capable of detecting small amounts of mirror asymmetry in two-dimensions, but find that this result does easily generalize to three dimensions. In a follow-up paper, we investigate deep scattering networks and neural fields and show they outperform the CNNs presented in this work. In the future we will explore architectures which are known to perform well on sparse three dimensional data~\cite{wu2019pointconv}, and point clouds~\cite{qi2017pointnet, qi2017pointnet++, qian2022pointnext, zhou2020graph} which have already been shown to perform well in the context of galaxy clustering~\cite{Anagnostidis:2022rbs, Villanueva-Domingo:2022rvn, Makinen:2022jsc}.

\par Compared to 4PCFs, the unsupervised approached can be trained from the data alone and is not limited by the number and accuracy of available simulations, nor does the detection significance rely on an accurate model of 8-point convariance contributions. The method is also in principle capable of extracting information from the higher-order $N$-point configurations, which quickly becomes computationally infeasible for $N$-point correlation functions at higher order. Furthermore one can make a map of $f$ over the survey volume to determine whether the parity-violating signal is spatially correlated with known systematics.
\par Despite these advantage, the unsupervised approach has several drawbacks. First, it is not possible to fit a parity-violating physical model to constrain parameters of interest. Second, as with many machine learning algorithms the method suffers from a lack of interpretability. For these reasons, the unsupervised and 4PCF approach should be thought of as complementary. 
\par The next step is to find an architecture which is capable of detecting parity-violations in simulations where we know the signal is present e.g.~\cite{Coulton:2023oug,Slepian:2023gni}. We will then search for parity-violations in the CMASS sample to verify the results of Ph21, Ph22 and Hou22. A detection would provide the first confirmation using an independent technique. 
\par The unsupervised approach presented in this work may become a powerful tool in the search for parity-violations in the large-scale structure of the Universe. The next generation of spectroscopic surveys will prove an important testing ground. Euclid, Roman and DESI will employ very different survey strategies, so parity-violating observational systematics are unlikely to be shared across all three datasets. Thus, confirmation of parity-violations using the methods presented in this paper in all three next generation datasets would put the existence of physical parity-violations on firmer ground.

\section{Acknowledgements}
 PLT is supported in part by NASA ROSES 21-ATP21-0050. This work received support from the U.S. Department of Energy under contract number DE-SC0011726. Y.S.T. acknowledges financial support from the Australian Research Council through DECRA Fellowship DE220101520. The authors thank Ashley Ross, Chun-Hao To, David Weinberg, Eric Huff, Zachary Slepian, Oliver Philcox, Erik Zaborowski, Maja Joblonska, Eduardo Rozo, Tomasz R\'{o}\.{z}a\'{n}ski, Iona Ciuc\u{a}, Biprateep Dey, Jiamin Hou and Robert Cahn for enlightening conversations. This work used public software~\cite{2020SciPy-NMeth, 2020NumPy-Array, 4160265}. This research used resources of the National Energy Research Scientific Computing Center (NERSC), a U.S. Department of Energy Office of Science User Facility located at Lawrence Berkeley National Laboratory, operated under Contract No. DE-AC02-05CH11231.

\bibliographystyle{apsrev4-1.bst}
\bibliography{bibtex.bib}

\appendix
\section{Network Architectures}
The network architectures used in Sect.~\ref{sec:results} are summarized in this Section. The networks are optimized up to the point where they are capable of detecting parity-violations, but we make no claims that these architectures are optimal. 

\subsection{2D Model 1 Architecture} \label{sec:A1}
For $g$, we take a fairly common network architecture using {\tt ReLu}~\cite{agarap2018deep} activation functions with a single convolution layers of $10$ filters with filter size of $(10,10)$ pixels, followed by a max pool layer of size $(2,2)$ with no strides. This is connected to a dense layer with $10$ neurons before the final layer outputs a single value, $g(x)$. The network is trained for $10$ epochs, using the Adam optimizer~\cite{kingma2014adam} with a constant learning rate of $10^{-3}$ and a batch size of $64$. The learning rate and batch size are fixed for every network in this paper.

\subsection{2D Model 2 Architecture} \label{sec:A2}
The network architecture used in this example is loosely based on AlexNet~\cite{krizhevsky2017imagenet}. The network starts with a convolution layer of 100 filters of size $(11,11)$ padded so that input and output size remains unchanged, followed by a max pool layer of size $(2,2)$ with no strides. This is followed by two convolution layers with 100 filters of size $(5,5)$ padded so that input and output size remains unchanged, then three convolution layers with 100 filters of size $(3,3)$, followed by a max pool layer of size $(2,2)$ with no strides. This is then sequentially passed to two dense layers with 4000 neurons with a $50 \%$ dropout rate before being passed to a dense layer with 1000 neurons and before outputting $g(x)$. {\tt ReLu} activation functions are used throughout. The network is trained for 10 epochs.

\subsection{3D Model 1 Architecture} \label{sec:A4}
The network starts with a 3D convolution layer with 100 filters of size 11 padded so that the size of the training data remains the same. This is followed by a max pool layer of size 2 with a stride length of 1. This is followed by four 3D convolutions of 100 filter of size 3 before being passed to a max pool layers 2 with a stride length of 1. After flattening, this is followed by 3 dense layers of size 4000, 1000 and 200 with a $50 \%$ dropout rate between each before outputting $g(x)$. {\tt ReLu} activation functions are used throughout. The network is trained for 10 epochs.

\end{document}